\documentclass[a4paper,11pt]{article}
\usepackage{pos}

\title{ Short \& intermediate distance HVP contributions to muon g-2: SM (lattice) prediction versus e+e- annihilation data}
\ShortTitle{Lattice windows for $g_\mu-2$ HVP}

\author[a,b]{C.~Alexandrou}
\author[b]{S.~Bacchio}
\author[c]{P.~Dimopoulos}
\author[b]{J.~Finkenrath}
\author*[d]{R.~Frezzotti}
\author[e]{G.~Gagliardi}
\author[f]{M.~Garofalo}
\author[a,b]{K.~Hadjiyiannakou}
\author[g]{B.~Kostrzewa}
\author[h]{K.~Jansen}
\author[e,j]{V.~Lubicz}
\author[f]{M.~Petschlies}
\author[e]{F.~Sanfilippo}
\author[e]{S.~Simula}
\author[f]{C.~Urbach}
\author[k]{U.~Wenger}

\affiliation[a]{Department of Physics, University of Cyprus, 20537 Nicosia, Cyprus}

\affiliation[b]{Computation-based Science and Technology Research Center, The Cyprus Institute,\\
20 Konstantinou Kavafi Street, 2121 Nicosia, Cyprus}

\affiliation[c]{Dipartimento  di  Scienze  Matematiche,  Fisiche  e  Informatiche,  Universit\`a  di  Parma  and  INFN -- Parma,\\ Parco  Area  delle  Scienze  7/a  (Campus),  43124  Parma,  Italy}

\affiliation[d]{Dipartimento di Fisica, Universit\`a di Roma ``Tor Vergata'' 
and INFN, Sezione di Roma Tor Vergata\\
Via della Ricerca Scientifica 1, I-00133 Rome, Italy}

\affiliation[e]{Istituto Nazionale di Fisica Nucleare, Sezione di Roma Tre,\\
Via della Vasca Navale 84, I-00146 Rome, Italy}

\affiliation[f]{HISKP (Theory), Rheinische Friedrich-Wilhelms-Universit\"at Bonn,\\
Nussallee 14-16, 53115 Bonn, Germany}

\affiliation[g]{High Performance Computing and Analytics Lab, Rheinische Friedrich-Wilhelms-Universit\"at Bonn,\\
Friedrich-Hirzebruch-Allee 8, 53115 Bonn, Germany}

\affiliation[h]{NIC, DESY, Platanenallee 6, D-15738 Zeuthen, Germany}

\affiliation[j]{Dipartimento di Matematica e Fisica, Universit\`a Roma Tre,\\
Via della Vasca Navale 84, I-00146 Rome, Italy}

\affiliation[k]{Institute for Theoretical Physics, Albert Einstein Center for Fundamental Physics,\\
University of Bern, Sidlerstrasse 5, CH-3012 Bern, Switzerland}

\emailAdd{roberto.frezzotti@roma2.infn.it}

\abstract{We present new lattice results of the ETM Collaboration, obtained from extensive simulations of lattice QCD with dynamical up, down, strange and charm quarks at physical mass values, different volumes and lattice spacings, concerning the SM prediction for the so-called intermediate window (W) and short-distance (SD) contributions to the leading order hadronic vacuum polarization (LO-HVP) term of the muon anomalous magnetic moment, $a_\mu$. 
Results for $a_{\mu}^{\rm LO-HVP,W}$ and $a_{\mu}^{\rm LO-HVP,SD}$, besides representing a step forward to a complete lattice computation of $a_{\mu}^{\rm LO-HVP}$ and a useful benchmark among lattice groups, are compared here with their dispersive counterparts based on experimental data for $e^+e^-$ into hadrons. The comparison confirms the tension in $a_{\mu}^{\rm LO-HVP,W}$, already noted in 2020 by the BMW Collaboration, while showing no tension in $a_{\mu}^{\rm LO-HVP,SD}$.}

\FullConference{%
  41st International Conference on High Energy physics - ICHEP2022\\
  6-13 July, 2022\\
  Bologna, Italy
}

\newcommand{\be}{\begin{equation}}
\newcommand{\ee}{\end{equation}}
\newcommand{\bea}{\begin{eqnarray}}
\newcommand{\eea}{\end{eqnarray}}

\begin{document}
\maketitle

\section{From the muon $g-2$ to probing the $R$-ratio of $e^+e^- \rightarrow$~hadrons}

The anomalous magnetic moment of the muon $a_{\mu} \equiv (g-2)/2$, one of the most accurately known quantities in physics, is a crucial observable for which a long-standing tension between the experimental value and the Standard Model (SM) prediction can provide evidence for New Physics (NP) beyond the SM. 
The current experimental world average\,\cite{Muong-2:2021ojo} is $a_{\mu}^{\rm exp} = 116\,592\,061 (41) \cdot 10^{-11}$, with a relative error of $0.35$ ppm.
Ongoing Fermilab data analyses should reduce the error by a factor of about four, and 
in future the E34 experiment at J-PARC will reach a similar precision.

On the theoretical side, the dominant source of uncertainty in the determination of $a_{\mu}$ comes from the leading-order Hadronic Vacuum Polarization (HVP) term $a_{\mu}^{\rm HVP}$ of order $\mathcal{O}(\alpha_{em}^{2})$, and to a less extent, from the Hadronic Light-by-Light (HLbL) scattering contribution of order $\mathcal{O}(\alpha_{em}^{3})$. The most precise prediction for the HVP contribution has come so far from a data-driven approach, where the result is reconstructed from the experimental cross section data for $e^+ e^-$ annihilation into hadrons, using dispersion relations plus a pure SM completion at high energy, and reads~\cite{Aoyama:2020ynm} 
\be
    \label{eq:HVP_dispersive}
    a_{\mu}^{\rm HVP, ddSM} = 6\,931 (40) \cdot 10^{-11} ~ ,  ~
\ee
where $40 \cdot 10^{-11}$ corresponds to an uncertainty on the full $a_{\mu}$ of $0.37$ ppm. 
The difference between the experimental result $a_{\mu}^{\rm exp}$ 
and the prediction of $a_\mu$, which is obtained using SM theory plus the dispersive result in Eq.~(\ref{eq:HVP_dispersive}) for the HVP term and is called the data-driven SM value $a_\mu^{\rm ddSM}$, amounts to\,\cite{Aoyama:2020ynm}
\be
    \label{eq:anomaly}
    \Delta a_{\mu} = a_{\mu}^{\rm exp} - a_{\mu}^{\rm ddSM} = 251 (41) (43) \cdot 10^{-11} = 251 (59) \cdot 10^{-11} ~ . ~
\ee
Here the first (second) error in the central expression comes from experiment (theory), while the total error is given in the last expression. The result~(\ref{eq:anomaly}) displays a remarkable $4.3 \sigma$ tension.

In order to check the data-driven SM prediction for $a_\mu^{\rm HVP}$, lattice field theory can play
a key role, as it allows to predict $a_\mu^{\rm HVP}$ from the pure SM 
theory, namely QCD+QED, renormalized in terms of $\alpha = 1/137.036...$ and few hadronic masses.
Within lattice QCD+QED $a_{\mu}^{\rm HVP}$ can be evaluated directly in the 
time-momentum representation\,\cite{Aoyama:2020ynm} as an integral over Euclidean time $t$ of the zero three-momentum Euclidean correlation function $V(t)$ (Eq.(\ref{eq:VV})) times the known function\footnote{The leptonic kernel $K(z)$ is proportional to $z^2$ at small values of $z$ and it approaches $1$ as $z \to \infty$.} $K(m_{\mu} t)$:
%
\be
    \label{eq:amu_HVP}
    a_{\mu}^{\rm{HVP}} = 2 \alpha_{em}^2 \int_0^\infty ~ dt \, t^2 \, K(m_\mu t) \,V(t) ~ , ~
\ee
\be
    \label{eq:kernel}
K(z) =  2 \int_0^1 dy ( 1- y) \left[ 1- j_0^2\left( zy / (2\sqrt{1 - y}\,) \right) \right]~, \qquad j_{0}(y) =
\sin(y)/y ~.~
\ee
The Euclidean vector correlator $V(t)$ can be calculated on a lattice with spatial volume $V = L^3$ and time extent $T$ for discretized values of the time distance $t / a$ from $0$ to $T / a$. It is defined as
 \be
     \label{eq:VV}
 V(t) \equiv - \frac{1}{3} \sum_{i=1,2,3} \int d^3{x} ~ \langle J_i(\vec{x}, t) J_i(0) \rangle \, ,
 \ee
with $J_\mu(x) \equiv \sum_{f = u, d, s, c, ...} q_f ~ \overline{\psi}_f(x) \gamma_\mu \psi_f(x)$
being the electromagnetic (em) current operator and $q_f$ the em charge for the quark flavor $f$ (in units of the absolute value of the electron charge).

A breakthrough in the accuracy for $a_\mu^{HVP}$ was obtained in the recent lattice SM calculation performed by the BMW Collaboration (BMW\,'20\,\cite{Borsanyi:2020mff}): $a_\mu^{\rm HVP,latSM}({\rm BMW}) = 7\,075 (55) \cdot 10^{-11}$, corresponding to a relative uncertainty of $0.8\%$. The result of the BMW Collaboration differs from the data-driven one\,(\ref{eq:HVP_dispersive}) at the level of $2.1\sigma$, thereby weakening the tension\,(\ref{eq:anomaly}) to a $1.5 \sigma$ effect. 

Further independent lattice SM determinations of $a_\mu^{HVP}$ with a few permille uncertainty are now required. This is a challenging task owing to the complexity of the computation if all sources of error are to be kept under control to such an high accuracy level.
In this respect, the so-called short and intermediate time-distance windows, introduced by the UKQCD-RBC Collaboration\,\cite{RBC:2018dos} represent important benchmark quantities. They are given by 
\be
    \label{eq:amu_w}
    a_\mu^{{\rm HVP}, w} = 2 \alpha_{em}^2 \int_0^\infty ~ dt \, t^2 \, K(m_\mu t) \, \Theta^w(t)\,V(t) ~ \qquad w = \{ SD, W, LD \} ~ , ~
\ee
where the time-modulating function $\Theta^w(t)$ reads
\be \label{eq:Mt_SDWLD}
  \Theta^{\rm SD}(t) \equiv  1 -  \frac{1}{1 + e^{- 2 (t - t_0) / \Delta}} ~ , 
  \quad
  \Theta^{\rm W}(t) \equiv \frac{1}{1 + e^{- 2 (t - t_0) / \Delta}} -  \frac{1}{1 + e^{- 2 (t - t_1) / \Delta}} ~ ,
  \quad
 \Theta^{\rm LD}(t) \equiv \frac{1}{1 + e^{- 2 (t - t_1) / \Delta}} ~ ,
\ee
with $t_0=0.4$~fm, $t_1=1$~fm, $\Delta = 0.15$~fm and 
$ a_\mu^{\rm HVP} \equiv a_\mu^{\rm HVP, SD} + a_\mu^{\rm HVP, W} + a_\mu^{\rm HVP, LD} $.
Indeed these ``window'' observables allow for comparisons not only among lattice results from different groups, but also between lattice results, i.e.\ {\it ab initio} SM predictions, and their data driven (``ddSM'') counterparts based on $e^+ e^-\rightarrow$~hadrons experiments. 

 \begin{figure}[htb!]
 \begin{center}
 
 \begin{minipage}{0.4\textwidth}
 \raggedleft
 \hspace{-2.8cm}
 \vspace{+0.2cm}
 \includegraphics[scale=0.29]{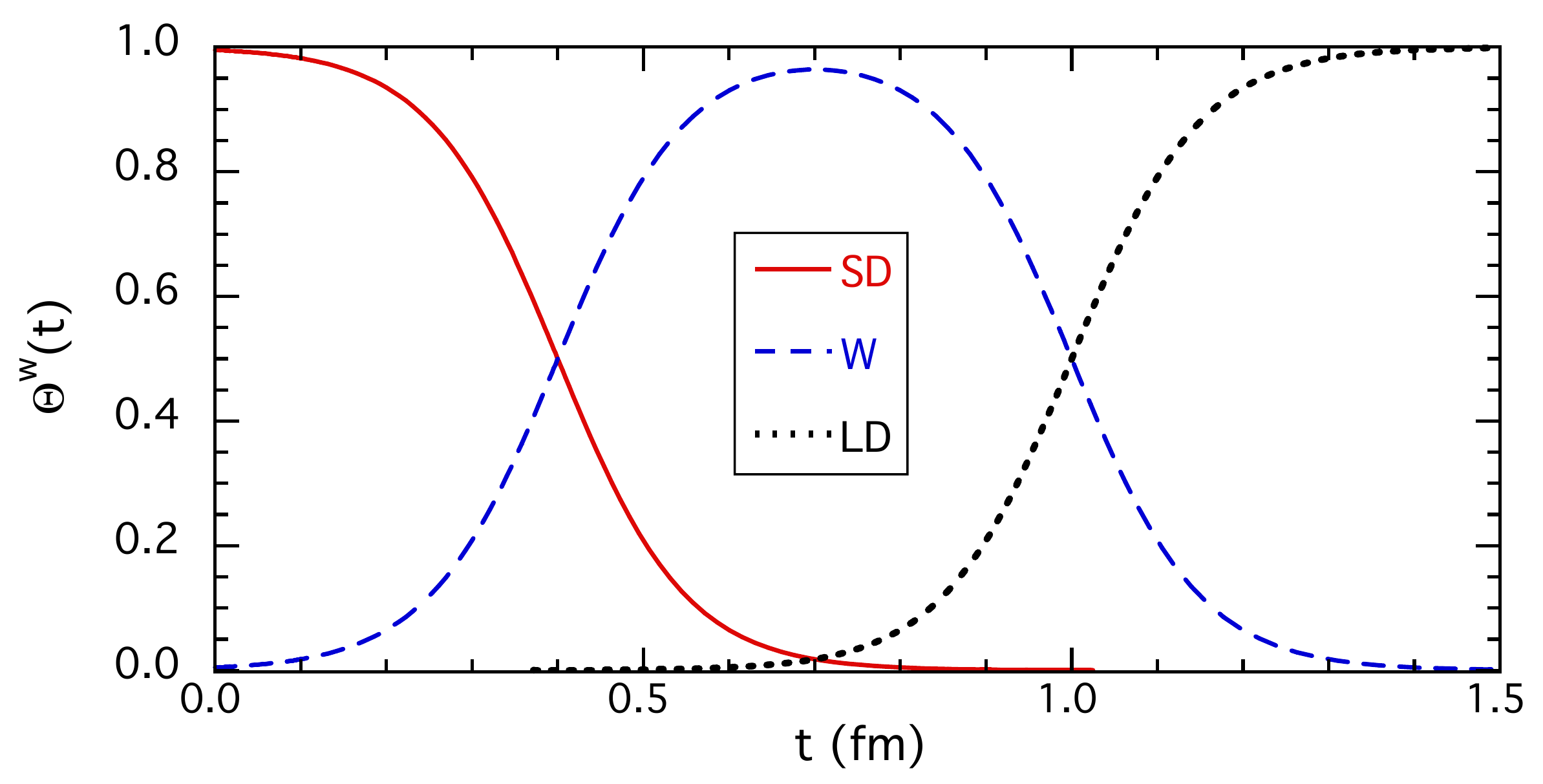}
 \end{minipage}
 \begin{minipage}{0.3\textwidth}
         \hspace*{1.0cm}
 \includegraphics[scale=0.23]{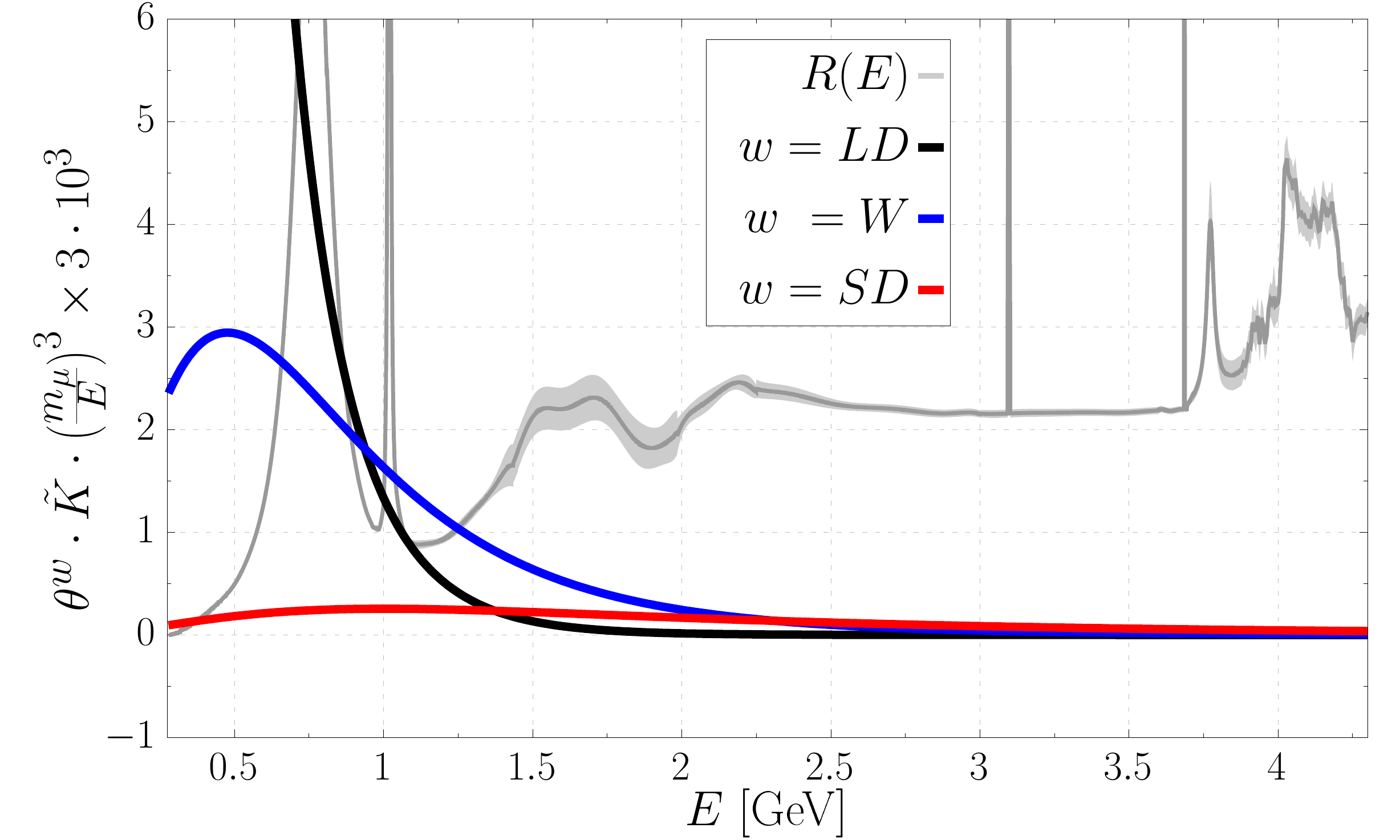}
 \end{minipage}
 \hspace*{1.5cm}
         \caption{\it \small Left panel: the function $\Theta^{w}(t)$ for $w=SD,\, W,\, LD$ defining $a_\mu^{\rm HVP,w}$, see Eq.\, (\ref{eq:amu_w}). Right panel: the weight
 $\frac{m_\mu^3}{E^3} ~ \widetilde{K}\left( \frac{E}{m_\mu} \right) \, \widetilde{\Theta}^w(E)$ and (overlayed in grey) the experimental $R^{had}(E)$, both appearing in Eq.\,(\ref{eq:amu_w_E}).}
 \label{fig:windows}
 \end{center}
 \end{figure}

\vspace*{-0.5cm}
The latter point becomes evident, see Eq.~(\ref{eq:amu_w_E}), upon rewriting $a_\mu^{\rm HVP, w}$ as an integral over the (center-of-mass) energy$E$ of the final hadron state in the $e^+e^-$ annihilation process with cross section
 \be
      \sigma^{had}(E) = \frac{4 \pi \alpha_{em}^2}{3 E^2} R^{had}(E) \, .
      \label{eq:Rhad}
 \ee
In fact, using the spectral representation $V(t) =  \frac{1}{12\pi^2} \int_{E_{thr}}^{\infty} dE E^2 R^{\rm had}(E) e^{-E t}$, one obtains
\be
      \label{eq:amu_w_E}
      a_\mu^{\rm HVP,w} = \frac{2 \alpha_{em}^2}{9 \pi^2 m_\mu} \, \int_{E_{thr}}^{\infty} dE \frac{m_\mu^3}{E^3} ~ \widetilde{K}\left( \frac{E}{m_\mu} \right) \, \widetilde{\Theta}^w(E) \, R^{had}(E) ~ , ~
\ee
where the energy-modulating function $\widetilde{\Theta}^w(E)$ and the 
leptonic kernel $\widetilde{K}(x)$ are given by\footnote{$\widetilde{K}(x)$ is proportional to $x^2$ for $x \ll 1$ and approaches $1$ as $x \to \infty$. At the two-pion threshold: $\widetilde{K}(2 M_\pi / m_\mu) \simeq 0.63$.}
\be
\widetilde{\Theta}^w(E)= \frac{\int_0^\infty dt ~ t^2 ~ e^{- E \, t} ~ K(m_\mu t) ~ \Theta^w(t)}{\int_0^\infty dt ~ t^2 ~ e^{- E \, t} ~ K(m_\mu t)} \; , \qquad
\widetilde{K}(x) = \frac{3}{4} x^5 \int_0^\infty dz ~ z^2 ~ e^{-x \, z} ~ K(z) ~.~
\ee
%
For $w= {\rm SD,  W, LD}$, the modulating functions $\Theta^{w}(t)$ and 
$\frac{m_\mu^3}{E^3} ~ \widetilde{K}\left( \frac{E}{m_\mu} \right) \, \widetilde{\Theta}^w(E)$
are shown in Fig.\,\ref{fig:windows}.

Here we present new accurate results for $a_\mu^{\rm HVP,W}$ and (for the first 
time) $a_\mu^{\rm HVP,SD}$, which can be directly compared with their data-driven counterparts
and represent an {\em ab initio} probe of the $R$-ratio
$R^{had}(E)$ weighted with the specific kernels
$\frac{m_\mu^3}{E^3} ~ \widetilde{K}\left( \frac{E}{m_\mu} \right) \, \widetilde{\Theta}^w(E)$, $w=W,SD$ (see~\cite{Alexandrou:2022amy} for details).

%

\section{Extended Twisted Mass Collaboration (ETMC) lattice data and other inputs}

We compute separately the $u$, $d$, $s$ and $c$ fermionic connected and 
disconnected contributions to the Euclidean correlator $V(t)$ (see Eq.(\ref{eq:VV})) 
and in terms of them we evaluate the window observables $a_\mu^{\rm HVP,SD}$ and 
$a_\mu^{\rm HVP,W}$ (see Eq. (\ref{eq:amu_w})). To this goal we exploit extensive
simulations of lattice QCD with dynamical $u$, $d$, $s$ and $c$ quark flavours in the 
isosymmetric limit ($\alpha_{em}= m_d-m_u = 0~~\Rightarrow~~u=d\equiv \ell$) -- here 
called "isoQCD" -- that have been presented in ETMC\,'22\,\cite{Alexandrou:2022amy} with 
\begin{itemize}
\item three (four in the case of $c$ contributions) lattice spacings used for continuum extrapolation;

%
\item accurate tuning of $s$ and $c$, besides $\ell$, quark masses in both valence and
sea fermion sectors;

\item O($10^3$) measurements on hundreds of gauge configurations for the
$\ell$ quark contributions;

\item vector currents with very precise (0.1\%) chiral covariant
normalization (hadronic method);

\item no dangerous O($a^2\,\log(a^2)$) artifacts in $a_\mu^{SD}$ (removed via direct 
tree-level computation).

\item physical pion mass\footnote{Recently evaluated corrections of
	our observables from the originally simulated $M_\pi$ values ($\sim \! 140$ or
	$\sim \! 137$~MeV) to $M_\pi^{\rm isoQCD}=135.0$~MeV gave 
better sensitivity to lattice artifacts, leading us to try and 
combine a larger number of fits. } 
and large volume systems ($L^3 \times 2L$), with $L$ in the range $5.1$\,fm -- $7.6$\,fm; 
the continuum limit is taken 
 on data interpolated at $L_{\rm ref}=5.46$\,fm, then moved to $L \to \infty$.

\end{itemize}

\begin{figure}[htb!]
\begin{center}

\begin{minipage}{0.4\textwidth}
\hspace*{-0.8cm}
\vspace{+0.2cm}
\includegraphics[scale=0.235]{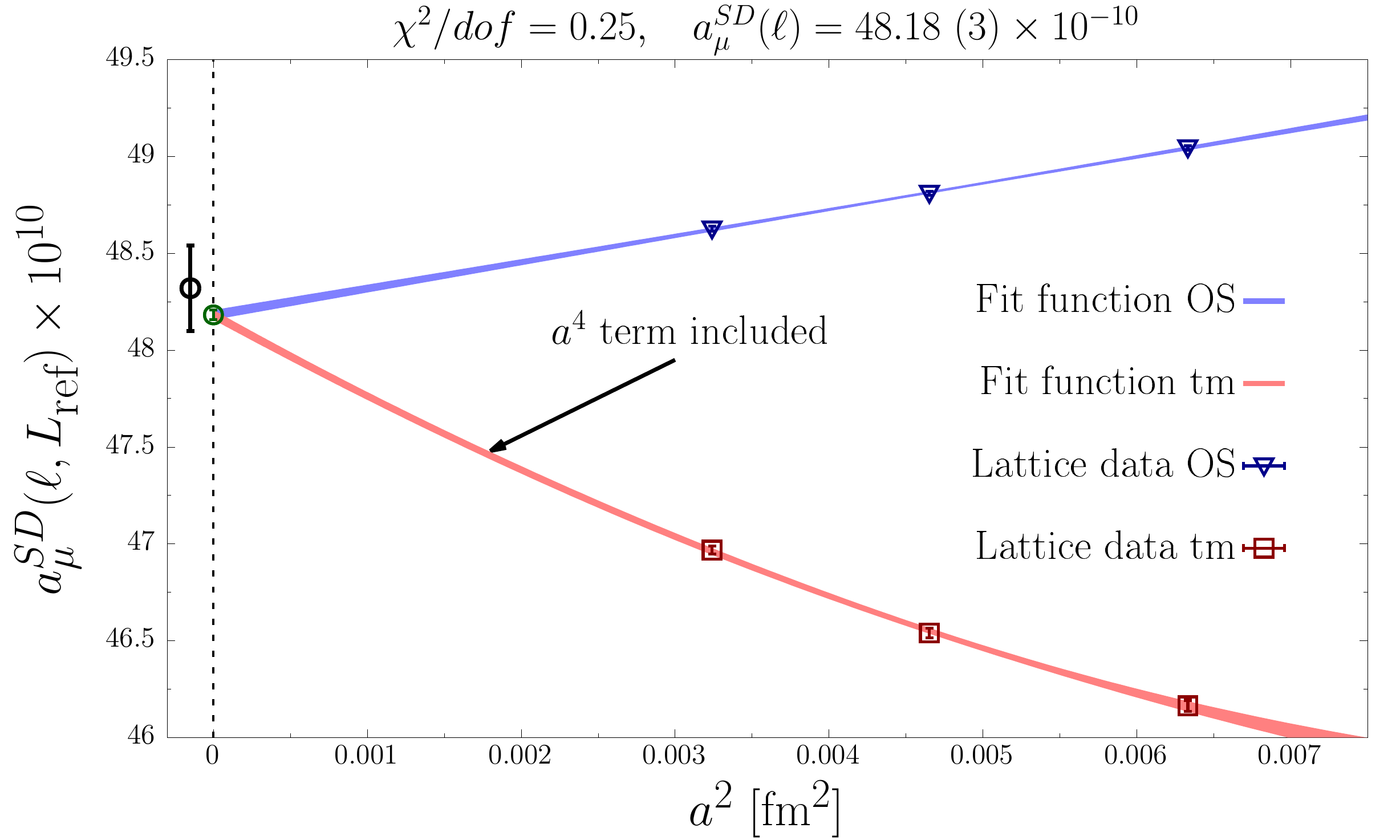}
\end{minipage}
\begin{minipage}{0.3\textwidth}
        \hspace*{0.2cm}
\includegraphics[scale=0.25]{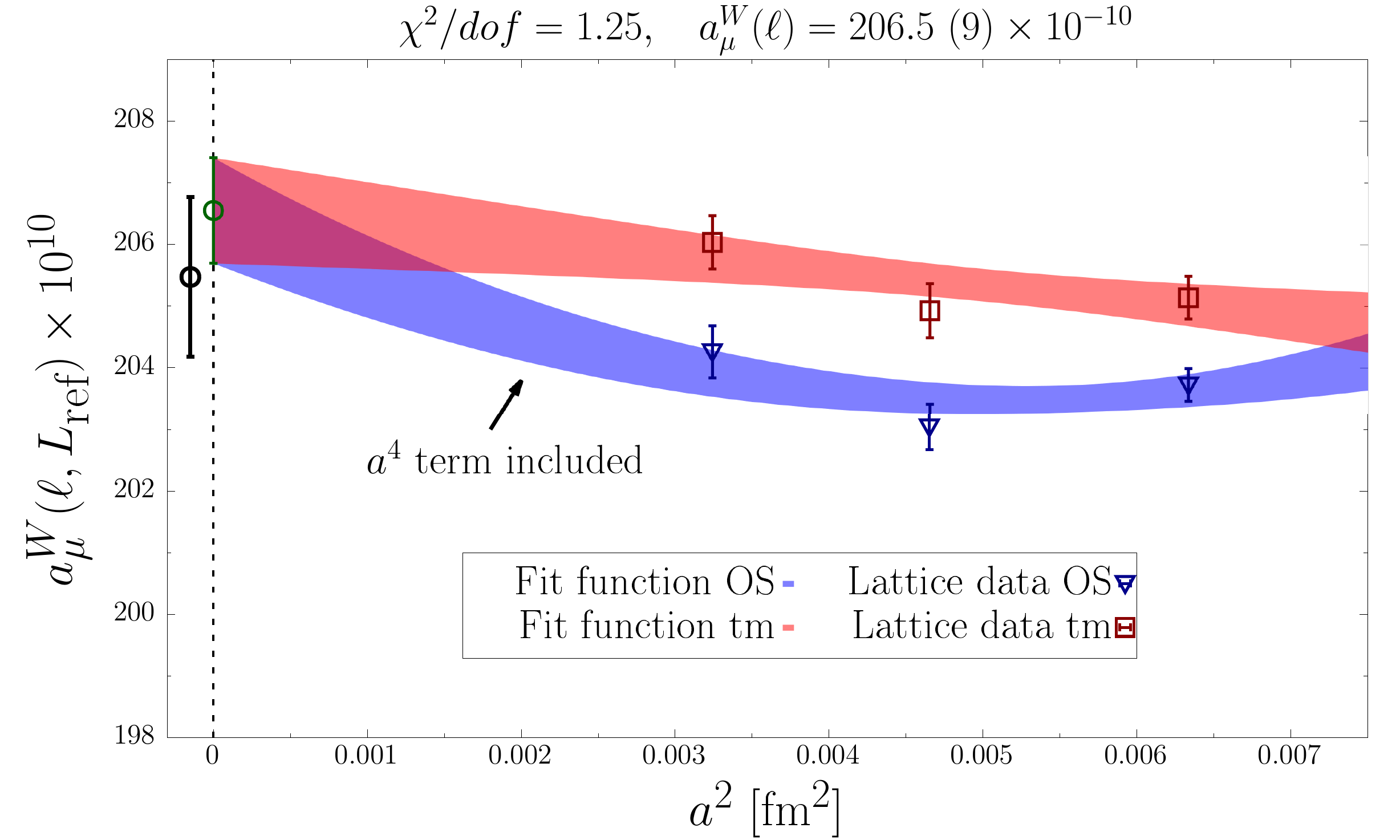}
\end{minipage}
\hspace*{1.5cm}
	\caption{\it \small Continuum extrapolation  
	of $a_\mu^{\rm HVP,SD}(\ell)$ (left) and $a_\mu^{\rm HVP,W}(\ell)$ (right) 
	data in two lattice regularizations (``tm'' and ``OS''), for
	$M_\pi^{\rm isoQCD} = 135.0$~MeV and the reference size $L_{\rm ref} = 5.46$~fm. 
	Legend info and coloured $1\sigma$ bands refer to one representative fit among
	the many that we considered. A black symbol, close to the dashed line, 
	shows the mean and total error for the combination of all fits.
	See~\cite{Alexandrou:2022amy} for analysis details.}
\label{fig:quality}
\end{center}
\end{figure}

\noindent An example of the data quality and the accuracy of the continuum extrapolation is shown
in Fig.~\ref{fig:quality}.

\noindent We also use few tiny and relatively accurate inputs not coming from ETMC\,'22 simulations, namely \\ 
i) QED and strong isospin breaking effects on $a_\mu^{\rm HVP,W}$ evaluated by BMW\,'20\,\cite{Borsanyi:2020mff}: $$ a_\mu^{\rm HVP,W}({\rm QED+SIB}) = 0.43(4) \cdot 10^{-10}~;$$ 
ii) $b$ quark and QED effects on $a_\mu^{\rm HVP,SD}$ estimated in perturbative QCD via the ``rhad'' package\,\cite{Harlander:2002ur}: $$a_\mu^{\rm HVP,SD}(b) = 0.32 \cdot 10^{-10}~, \qquad a_\mu^{\rm HVP,SD}({\rm QED}) = 0.03 \cdot 10^{-10}~.$$


\section{Lattice SM results and comparison with data-driven determinations}

\begin{figure}[htb!]
\begin{center}
\includegraphics[scale=0.69]{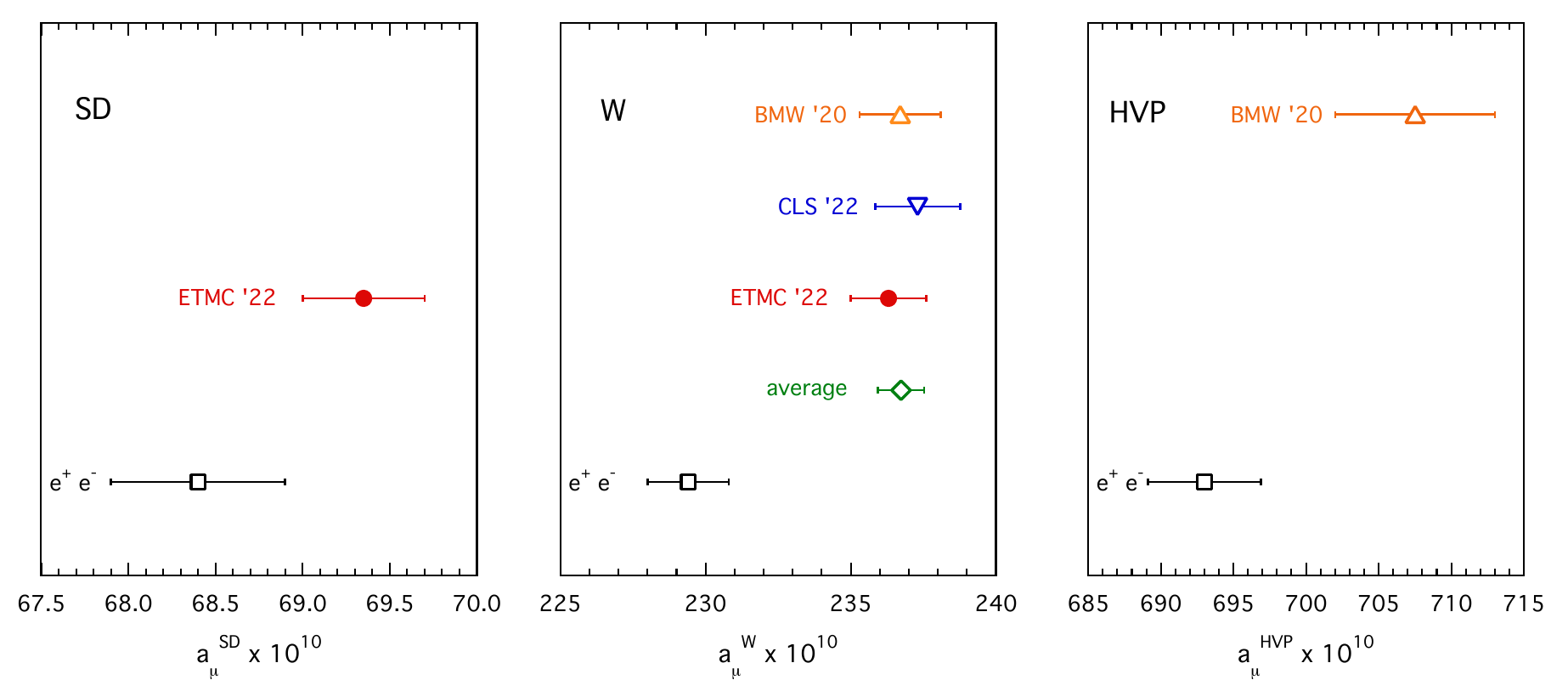}
\caption{\it \small Lattice SM results for the $a_\mu^{\rm HVP, SD}$ (left panel), 
	$a_\mu^{\rm HVP, W}$ (central panel) and full $a_\mu^{\rm HVP}$ (right panel)
observables, compared with their experimental data-driven counterparts~\cite{Colangelo:2022vok}. 
	Only results from at least three lattice spacings and one ensemble at the physical pion mass point are considered. Central panel: the green diamond is our average of the BMW\,'20, CLS\,'22 and ETMC\,'22 results: $~a_\mu^{\rm HVP, W} \; = \; 236.7(8) \cdot 10^{-10}~$.}
\label{fig:comparison}
\end{center}
\end{figure}

Our current (almost final) results, accounting for info from recent simulations at $M_\pi=135$~MeV and 
for an analysis with an enlarged set of fits combined in different ways, may be summarized as follows.
For the observable $a_\mu^{\rm HVP,W}$, probing the $R$-ratio at low and intermediate $E$, we obtain
\bea
&&  a_\mu^{\rm W}(\ell, \; s, \; c, \; {\rm disc}) = 
[206.5(1.3), \; 27.28(20) , \; 2.90(12), \; -0.78(21) ] \cdot 10^{-10} \, ,  \\
&& {\rm yielding} \qquad a_\mu^{\rm W}({\rm ETMC}) \; = \; 236.3 (1.3) \cdot 10^{-10} \, .
\eea
The short distance observable $a_\mu^{\rm HVP,W}$ probes the $R$-ratio at higher $E$
(see Fig.~\ref{fig:windows}). For it we find
\bea
&&  a_\mu^{\rm SD}(\ell, \; s, \; c, \; {\rm disc}) = 
[48.32(22), \; 9.074(64) , \; 11.61(27), \; -0.006(5) ] \cdot 10^{-10} \, ,  \\
&& {\rm yielding} \qquad a_\mu^{\rm SD}({\rm ETMC}) \; = \; 69.35 (35) \cdot 10^{-10} \, .
\eea
Our findings for partial flavour contributions to $a_\mu^{\rm HVP,W}$ are in remarkable 
agreement with those from other lattice groups (see~\cite{Alexandrou:2022amy} for details). 
Our ETMC\,'22 result for $a_\mu^{\rm HVP,W}$ agrees very well with its analog in the
BMW\,'20\,\cite{Borsanyi:2020mff} and CLS\,'22\,~\cite{Ce:2022kxy} papers.
A recent result for $a_\mu^{\rm HVP,SD} + a_\mu^{\rm HVP,W}$ from Fermilab
Lattice/HPQCD/MILC groups~\cite{FermilabLattice:2022smb} also confirms our findings.
So far only BMW\,'20\,\cite{Borsanyi:2020mff} has published a very precise, pure 
lattice-SM result on the (LO i.e.  O($\alpha_{em}^2$)) full $a_\mu^{\rm HVP}$, and only 
ETMC\,'22\,\cite{Alexandrou:2022amy} has computed $a_\mu^{\rm HVP,SD}$. 
A concise summary of the situation is given in Fig~\ref{fig:comparison}, where we also show 
a comparison with $e^+e^- \to$ hadrons data-driven determinations of the same quantities.

The self-consistency of all lattice results
enhances the credibility of the full $a_{\mu}^{\rm HVP}$ result by BMW\,'20. 
Our $a_{\mu}^{\rm HVP,W}$ lattice average in Fig.~\ref{fig:comparison} shows
a $4.5~\sigma$ tension with the $e^+e^- \to$~hadrons data-driven determination of 
Ref.~\cite{Colangelo:2022vok}, 
which adopts the conservative data merging procedure from Ref.~\cite{Aoyama:2020ynm}, and 
an even stronger one ($\simeq 6.1~\sigma$) with respect to the data-driven result of 
Ref.~\cite{Keshavarzi:2019abf}. This striking low energy anomaly in $a_{\mu}^{\rm HVP,W}$
definitely needs to be understood.

A good agreement (at $1.5~\sigma$ level) is instead 
seen between lattice and data-driven determinations 
of $a_{\mu}^{\rm HVP, SD}$, which probes the $R^{had}(E)$-ratio at higher $E$, where the 
photon HVP (i.e.\ $\Delta \alpha_{em}$) is indeed known (see~\cite{Ce:2022eix} and refs.\ 
therein) to be consistent with electroweak precision tests of the SM.

\acknowledgments
We thank all members of ETMC for the most enjoyable collaboration. We are very grateful to Guido Martinelli and Giancarlo Rossi for many discussions on the lattice setup and the methods employed in this work. We thank Nazario Tantalo for valuable discussions about the physical information that can be obtained by comparing experimental data on $e^+e^- \to$~hadrons with SM lattice predictions for observables related to the photon HVP term.


\vspace*{-0.1cm}
\begin{thebibliography}{99}

\bibitem{Muong-2:2021ojo}
B.~Abi \textit{et al.} [Muon g-2],
Phys. Rev. Lett. \textbf{126} (2021) no.14, 141801
doi:10.1103/PhysRevLett.126.141801
[arXiv:2104.03281 [hep-ex]].

\bibitem{Aoyama:2020ynm}
T.~Aoyama, \textit{et al.}
Phys. Rept. \textbf{887} (2020), 1-166
doi:10.1016/j.physrep.2020.07.006
[arXiv:2006.04822 [hep-ph]].


\bibitem{Borsanyi:2020mff}
S.~Borsanyi, \textit{et al.}
Nature \textbf{593} (2021) no.7857, 51-55
doi:10.1038/s41586-021-03418-1
[arXiv:2002.12347 [hep-lat]].

\bibitem{RBC:2018dos}
T.~Blum \textit{et al.} [RBC and UKQCD],
Phys. Rev. Lett. \textbf{121} (2018) no.2, 022003
doi:10.1103/PhysRevLett.121.022003
[arXiv:1801.07224 [hep-lat]].

\bibitem{Alexandrou:2022amy}
C.~Alexandrou, S.~Bacchio, P.~Dimopoulos, J.~Finkenrath, R.~Frezzotti, G.~Gagliardi, M.~Garofalo, K.~Hadjiyiannakou, B.~Kostrzewa and K.~Jansen, \textit{et al.}
[arXiv:2206.15084 [hep-lat]].

\bibitem{Harlander:2002ur}
R.~V.~Harlander and M.~Steinhauser,
Comput. Phys. Commun. \textbf{153} (2003), 244-274
doi:10.1016/S0010-4655(03)00204-2
[arXiv:hep-ph/0212294 [hep-ph]].


\bibitem{Ce:2022kxy}
M.~C\`e, A.~G\'erardin, G.~von Hippel, R.~J.~Hudspith, S.~Kuberski, H.~B.~Meyer, K.~Miura, D.~Mohler, K.~Ottnad, P.~Srijit, \textit{et al.}
[arXiv:2206.06582 [hep-lat]].

\bibitem{FermilabLattice:2022smb}
C.~T.~H.~Davies \textit{et al.} [Fermilab Lattice, HPQCD and MILC],
[arXiv:2207.04765 [hep-lat]].

\bibitem{Colangelo:2022vok}
G.~Colangelo, \textit{et al.}
Phys. Lett. B \textbf{833} (2022), 137313
doi:10.1016/j.physletb.2022.137313
[arXiv:2205.12963 [hep-ph]].

\bibitem{Keshavarzi:2019abf}
A.~Keshavarzi, D.~Nomura and T.~Teubner,
Phys. Rev. D \textbf{101} (2020) no.1, 014029
doi:10.1103/PhysRevD.101.014029
[arXiv:1911.00367 [hep-ph]]; \,\,private commun.\, 2022.

\bibitem{Ce:2022eix}
M.~C\`e, A.~G\'erardin, G.~von Hippel, H.~B.~Meyer, K.~Miura, K.~Ottnad, A.~Risch, T.~San Jos\'e, J.~Wilhelm and H.~Wittig,
JHEP \textbf{08} (2022), 220
doi:10.1007/JHEP08(2022)220
[arXiv:2203.08676 [hep-lat]].

\end{thebibliography}
\end{document}